\documentclass[pra,twocolumn]{revtex4-1}
\usepackage{graphicx}
\usepackage{dcolumn}
\usepackage{bm}
\usepackage [latin1]{inputenc}
\usepackage{epstopdf}
\usepackage{color}
\usepackage{ulem}
\usepackage{multirow}
\begin{document}

\title{{Vortex generation in  a superfluid gas of dipolar chains in crossed
electric and magnetic fields}}

\author{D.V.Fil$^{1,2}$, S.I.Shevchenko$^3$}

\email{fil@isc.kharkov.ua}

\affiliation{$^1$ Institute for Single Crystals, National Academy of Sciences of Ukraine,
60 Nauky Avenue, Kharkiv 61072, Ukraine\\
$^2$V.N. Karazin Kharkiv National University,
4 Svobody Square, Kharkiv 61022, Ukraine\\
$^3$B.~Verkin Institute for Low Temperature Physics and Engineering, National Academy of
Sciences of Ukraine, Lenin av. 47 Kharkov 61103, Ukraine}

\begin{abstract}

 Crossed electric and magnetic fields influence dipolar neutral
particles in the same way as the magnetic field influences charged particles. The effect
of crossed fields is proportional to the dipole moment of the particle (inherent or
induced). We show that this effect  is quite spectacular in a multilayer system of polar
molecules.  In this system molecules  may bind in chains. At low temperature the gas of
chains becomes the superfluid one. The crossed fields then induce vortices in the
superfluid gas of chains. The density of vortices is proportional to the number of
particles in the chain. The effect can be used for monitoring the formation and
destruction of chains in multilayer dipolar gases.
   \end{abstract}

\maketitle

Neutral particles in crossed electric and magnetic fields behave as if they are charged
and subjected to an effective magnetic field. The effective magnetic field can be
expressed through an effective vector potential which is proportional to the vector
product of the magnetic field $\mathbf{B}$ and the dipole moment $\mathbf{d}$ of the
particle. The direction of the dipole moment can be fixed by the external electric field.
To produce nonzero effective vector potential the electric and magnetic fields should be
noncollinear.

The effective vector potential applied to a neutral superfluid can induce a superfluid
current \cite{1,4}. At least one the fields (the electric or  magnetic one) should be
nonuniform. Otherwise, the effective vector potential can be eliminated by the gauge
transformation. The effective vector potential and the effective magnetic field applied
to a dipole particle can be described in terms of the Aharonov-Bohm phase \cite{2,3}. Two
spatially separated positive and negative charges of a dipole particle feel slightly
different vector potentials and acquire slightly different in module and opposite in sign
phases under the motion of the dipole particle.  The overall phase is equal to the flux
of the magnetic field through the area  covered by the vector $\mathbf{r}_d=\mathbf{d}/e$
during its motion. One can introduce an effective vector potential and express the
Aharonov-Bohm phase through the integral of this potential. The effective magnetic field
is defined as the curl of the effective vector potential.

Generation of vortices in electrically neutral superfluids by a nonuniform magnetic field
was considered in \cite{5,6} (see also \cite{my} for a review) with reference to a
superfluid gas of electron-hole pairs in a bilayer system. The bilayer system consists of
two conducting layers with carriers of opposite signs separated by a dielectric layer.
The dipole moment of an electron-hole pair in a bilayer is proportional to the distance
between the layers and it can reach the value up to 10$^3$ Debyes. Therefore moderate
magnetic fields are required to generate quantum vortices. Generation of vortexes in
bilayers in a nonuniform electric field and uniform magnetic field was considered in
\cite{7}. Due to large polarizability of electron-hole pairs vortices are generated
already in moderate electric fields.

The typical value of the dipole moment of a polar molecule  does not exceed 5 Debyes
\cite{8,9,9a} and much larger magnetic fields are required to generate vortices. Dipolar
gases placed into a multilayer trap demonstrate a tendency to form chains
\cite{15,18,z1}. The phenomenon is connected with that the dipole-dipole interaction
between molecules located at different layers is attractive at small distances. The
dipole moment of a chain is proportional to the number of molecules in a chain and it is
comparable in value to the dipole moment of the electron-hole pair in the bilayer.

It is expected that increasing of the dipole strength results in a phase transition from
a molecular superfluid to a dipole chain superfluid \cite{15}, and then in a transition
from the chain superfluid  to a dipole Wigner crystal \cite{w1}. In a system of Fermi
polar molecules in a multilayer trap a transition from the dipole chain superfluid to a
dimerized superfluid may take place \cite{17}. The vortices generated by the crossed
fields   should disappear or their density should reduce considerably under the
transitions from a chain superfluid to a molecular or dimerized superfluid.  Vortexes
also disappear in the Wigner crystal state.

Let us derive the effective vector potential and the effective magnetic field using the
conception of the Aharonov-Bohm phase. The motion of a dipole particle along a closed
loop results in generation of two Aharonov-Bohm phases, one is for the positive charge
and the other, for the negative one. The sum of two phases is  \begin{equation}\label{1}
    \varphi_{A-B,d}=\frac{e}{\hbar c}\left(\oint_{C_+} \mathbf{A}d
    \mathbf{r}-\oint_{C_-} \mathbf{A}d
    \mathbf{r}\right)=\frac{e}{\hbar c}\int_{\Delta S} \mathbf{B} d
    \mathbf{S},
\end{equation}
where $C_+$ and $C_-$ are the paths of the positive and negative charges,
correspondingly, and  $\Delta S$ is the
 area covered by the vector $\mathbf{r}_d$ under such a motion.
In a typical situation a variation of the magnetic field and  the dipole moment is small
at the scale of $r_d$. Then the phase (\ref{1}) can be expressed through the integral of
the effective vector potential $\mathbf{A}_{eff}$ along the center of mass path $C$:
\begin{equation}\label{2}
\varphi_{A-B,d}=\frac{e}{\hbar c}\oint_{C} \mathbf{A}_{eff}\cdot d \mathbf{R},
\end{equation}
where
\begin{equation}\label{3}
\mathbf{A}_{eff}=\frac{\mathbf{B}\times \mathbf{d}}{e}.
\end{equation}
Introducing the effective magnetic field
\begin{equation}\label{3a}
\mathbf{B}_{eff}=\nabla\times\mathbf{A}_{eff}
\end{equation}
one can rewrite the phase (\ref{2}) through the integral over the area surrounded by the
contour $C$:
\begin{equation}\label{2a}
\varphi_{A-B,d}=\frac{e}{\hbar c} \int_{S_C}\mathbf{B}_{eff}\cdot d \mathbf{S}.
\end{equation}
Taking into account that the dipole moment of the particle or its direction may depend on
the coordinate we obtain the following expression for the effective magnetic field:
\begin{equation}\label{4}
   \mathbf{B}_{eff}=\frac{1}{e}
   \left[\mathbf{B}(\nabla\cdot\mathbf{d})+(\mathbf{d}\cdot\nabla)\mathbf{B}-
(\mathbf{B}\cdot\nabla)\mathbf{d}\right].
\end{equation}
The effective field (\ref{4}) influences   neutral dipolar particles in the same way as
the real magnetic field influences   charged particles with the charge $+e$.

There is an essential difference between the Aharonov-Bohm effect for a charged particle
and for a dipole particle. For a charged particle the Aharonov-Bohn phase can be nonzero
even if the particle moves in the space where the magnetic field is zero. In contrast
the real magnetic field  should be nonzero along the path $C$ to produce the
Aharonov-Bohn phase.

If the dipole moment does not depend on $\mathbf{R}$ the component of the effective field
parallel to $\mathbf{d}$ can be expressed though the two-dimensional divergence of the
magnetic field:
\begin{equation}\label{5}
   {B}_{z,eff}=\frac{d}{e}\partial_z B_z=-\frac{d}{e}\left(\partial_x B_x+ \partial_y
   B_y\right).
\end{equation}
Here the $z$-axis is directed along $\mathbf{d}$.

The superfluid chain phase emerges in a strong electric field directed perpendicular to a
stack of two-dimensional (2D) traps. The dipole moment of a chain is directed along the
electric field and does not depend on the coordinate. To produce the effective magnetic
field the real magnetic field should have nonzero component in the plane parallel to 2D
traps and be nonuniform.

Here we consider two configurations of the magnetic field. The first one emerges at the
face end of a solenoid. The second is generated by a flat coil.
 The magnetic field of a solenoid is circularly symmetric. Near the face end it has nonzero radial component. The projection of the magnetic field to the plane of the face end of a long solenoid is equal to
$\mathbf{B}_{pl}\approx B_s\mathbf{r}/4 R_s$, where  $B_s$ is the magnetic field deep
inside the solenoid,  $R_s$ is the solenoid radius, and $r$ is counted from the solenoid
axis  (the inequality $r<R_s$ is implied).
The normal to the face end plane component of the effective magnetic field is uniform:
\begin{equation}\label{6a}
   {B}_{z,eff}=-\frac{d}{2 e} \frac{B_s}{ R_s}.
\end{equation}
A flat coil induces  the radial magnetic field in the coil plane. For the distance $a$
from the coil  much smaller than the coil radius $R_{coil}$, and for $a<r<R_{coil}$ ($r$
is counted from the coil axis) the magnetic field can be approximated as
$\mathbf{B}_{pl}\approx H_{coil} \mathbf{r}/r$, where $H_{coil}=2\pi I n_{coil}$, $I$ is
the electrical current in the coil and $n_{coil}$ is the  density of turns of the coil.
The effective magnetic field normal  to the coil plane  is nonuniform:
\begin{equation}\label{7}
   {B}_{z,eff}(r)=-\frac{d}{e} \frac{H_{coil}}{r}.
\end{equation}

 The crossed fields generate vortices if the effective flux $\Phi_{eff}$ (the flux of the effective magnetic field) through the Bose cloud exceeds the critical value  $\Phi_c$. This critical value depends on the particle density profile in the trap and the dependence of ${B}_{z,eff}$ on coordinate.
 We specify the case of a  axially symmetric multilayer harmonic trap centered at $r=0$ with layers parallel to the face end plane or the coil plane.
Then
\begin{equation}\label{7a}
    \Phi_{c}=f\Phi_0
    \left(\ln\frac{R_{TF}}{\xi_0}-\frac{1}{2}\right),
\end{equation}
where $\Phi_0=2\pi \hbar c/e$ is the flux quantum, $R_{TF}$ is the Tomas-Fermi radius of
a Bose cloud, $\xi_0$ is the vortex core radius ($\xi_0\ll R_{TF}$), and $f$ is the
numerical factor equal to $f=2$ for the case of the uniform effective
 field (\ref{6a}), and  to $f=3/4$ for the field (\ref{7}).
The effective flux $\Phi_{eff}$ can be expressed through the real magnetic field $B$ at
$r=R_{TF}$:
\begin{equation}\label{11}
{\Phi_{eff}}=\frac{2 \pi R_{TF} d }{e}B_{pl}(R_{TF})=\frac{d R_{TF}}{e \ell^2}{\Phi_0},
\end{equation}
where $\ell=\sqrt{\hbar c/e B_{pl}(R_{TF})}$ is the magnetic length. Taking
$B_{pl}(R_{TF})=0.1$ T, $R_{TF}=500$ $\mu$m and $d=3.5$ Debye we obtain
${\Phi_{eff}}\approx 6{\Phi_0}$.  It corresponds to a state with one or few vortexes.

Let us now consider the conditions of emergence of a superfluid chain phase in a stack of
2D  traps. We imply the same density of particles in each trap and equal distances $b$
between next neighbour 2D traps. The electric field aligns the dipole moments normally to
the 2D traps. The interaction between dipoles located in the same ($n=0$) or different
($n\ne 0$) traps is given by equation
\begin{equation}\label{14v}
   V_n(r)=\frac{d^2[r^2-2 (n b)^2]}{[r^2+(n b)^2]^{5/2}},
\end{equation}
where $r$ is the 2D radius vector, and $n$ is the distance between the traps in units of
$b$. Since the dipole-dipole interaction is attractive for the molecules located in
different layers not far from each other it may cause binding of molecules from different
layers.

The interaction strength is characterized by the dimensionless parameter
\begin{equation}\label{14}
    U_0=\frac{d^2 m}{\hbar^2 b},
\end{equation}
where $m$ is the mass of the molecule. In two dimensions a particle in the potential
$\lambda V(x)$ that satisfies the conditions  $\int d^2 x V(x)=0$ and $V(\infty)=0$ has a
bound state at any $\lambda$ \cite{d1} (see, also \cite{10}). The potential (\ref{14v})
is of that form. Two polar molecules from the adjacent layers bind in a pair at any
$U_0$, but at small $U_0$ the binding energy is exponentially small: $E_b\sim (\hbar^2/m
b^2)\exp(-8/U_0^2)$ \cite{d2}. At large $U_0$  the formation of chains can be described
analytically.

At large $U_0$ one can use the harmonic approximation for the potential (\ref{14v}):
\begin{equation}\label{12}
    V_n(r)\approx-\frac{2 d^2}{(n b)^3}+\frac{6 d^2 r^2}{(n b)^5}.
\end{equation}
 The energy of the bound state  of two particles in the potential (\ref{12}) with $n=1$ is equal to
\begin{equation}\label{15}
    E_{b,2}=-\frac{2 d^2}{b^3}\left(1-\sqrt{\frac{6}{U_0}}\right).
\end{equation}
The energy (\ref{15}) is the sum of the classical binding energy and the zero-point
energy of quantum fluctuations. The binding energy per molecule is  $E_{b,2}/2$.

Considering a long chain $N\gg 1$ and neglecting the edge effects we obtain the following
classical binding energy for the dipole chain:
\begin{equation}\label{15a}
E_{cl,N}=-\frac{2 N d^2}{b^3}\sum_{n=1}^\infty \frac{1}{n^3}=-\frac{2 N
d^2}{b^3}\zeta(3),
\end{equation}
where $\zeta(s)$ is the zeta-function ($\zeta(3)\approx 1.2$), and $N$ is the number of
molecules in the chain. The spectrum of low energy excitations of the chain contains two
degenerate transverse modes with the energies
\begin{equation}\label{15-a}
 \Omega(q)=\frac{2
d^2}{b^3}\sqrt{\frac{12}{U_0}}\sqrt{\sum_{n=1}^\infty \frac{\sin^2\left(\frac{q n
b}{2}\right)}{n^5}}.
\end{equation}
 The zero-point
energy is $E_{zp}=\sum_{q=0}^{2\pi/b}\Omega(q)$. The sum $E_{cl,N}+E_{zp}$ yields the
binding energy
\begin{equation}\label{18}
    E_{b,N}\approx-\frac{2
    N d^2}{b^3}\left(1.2-0.92\sqrt{\frac{6}{U_0}}\right).
\end{equation}
The binding energy should be negative. Therefore the harmonic approximation (\ref{12}) is
justified only at large $U_0$ ($U_0> 6$). For LiK molecules with  $d=3.5 $ Debye  and $m=
7.6\cdot 10^{-23}$ g for the stack with $b=250$ nm  the interaction strength parameter is
equal to $U_0\approx 30$.

One can see from (\ref{15}) and (\ref{18}) that for a long chain the binding energy per
molecule $E_{b,N}/N$ is more than in two times larger than of the same energy for a dimer
($E_{b,2}/2$). The edge effect  reduces the binding energy, and the binding energy per
molecule decreases under decrease of $N$. Therefore it is energetically preferable for
the molecules to bind in the longest chains ($N$-segment chains for the $N$-layer
system).

At large density the chains  overlap due to transverse vibrations. Overlapping may cause
destruction of the chains. To evaluate  the effect of vibration we calculate the average
square transverse displacement of molecules in the chain:
\begin{equation}\label{19}
   \langle\zeta^2\rangle ={2}\sum_{q\neq 0}\frac{\hbar^2}{2m N
    \Omega(q)}\left(1+2N_B[\Omega(q)]\right),
\end{equation}
where $N_B(\Omega)=(e^{\Omega/T}-1)^{-1}$ is the Bose distribution function. Calculation
of the integral over $q$ in (\ref{19}) yields
\begin{equation}\label{19-a}
    \langle\zeta^2\rangle\approx
    b^2\left(\frac{1}{\pi\sqrt{12 \zeta(3)
    U_0}}\ln N+\frac{1}{12\pi^2 U_0 \zeta(3)}\frac{T}{T_d}n_{ch} b^2\right),
\end{equation}
where $n_{ch}$ is the density of the chains, $T_d=\hbar^2 n_{ch}/M$ is the temperature of
degeneracy, and $M=N m$ is the mass of the chain.
 One can see  that for reasonable $N$
($N\lesssim 100$), $U_0\gg 1$, $T<T_d$ and $n_{ch} b^2\lesssim 1$ the average
displacement satisfies inequality $\langle\zeta^2\rangle\ll b^2$. Thus for the densities
of chains of order of $b^{-2}$ or smaller the condition $n_{ch}\zeta^2\ll 1$ is fulfilled
and overlapping of the chains  is small.

If polar molecules are bosons, the chains also satisfy the Bose statistics. The 2D gas of
such chains goes into the superfluid state. The transition into the superfluid state is
of the Berezinskii-Kosterlitz-Thouless type. The transition temperature $T_c$ is given by
equation
\begin{equation}\label{21}
    T_c=\frac{\pi}{2}\frac{\hbar^2}{M}n_s(T_c),
\end{equation}
where $n_s(T)$ is the superfluid density at finite temperature.

For the contact interaction between  particles the superfluid density is obtained from
the equation
\begin{equation}\label{21a}
\frac{n_{ch}-n_s}{n_{ch}}=\frac{3  \zeta(3)}{2 \pi}
\left(\frac{T}{T_d}\right)^3\left(\frac{\hbar^2}{M\gamma}\right)^2,
\end{equation}\label{21b}
where $\gamma$ is the constant of the contact interaction. For the chains the constant
$\gamma$ is evaluated as $\gamma\sim N d^2/W$ \cite{z2}, where $W$ is the width of the
individual trap. Since $\gamma\gg \hbar^2/M$, the difference between the superfluid
density $n_s$ and the total density $n_{ch}$ is small at $T<T_d$. Thus, with a good
accuracy the critical temperature is given by the expression $ T_c={\pi\hbar^2 n_{ch}}/{2
M}$.

Taking $n_{ch}= b^{-2}$, $b=250$ nm and $N=100$, we obtain the critical temperature
$T_c\approx 3$ nK for LiK molecules. For $B_{pl}(R_{TF})=0.1$ T and $R_{TF}=500$ $\mu$m
the effective flux reaches the value of ${\Phi_{eff}}\approx 6\cdot 10^2{\Phi_0}$. It
corresponds to a multivortex state with the average density of vortices $n_v\approx
8\cdot 10^4$ cm$^{-2}$. For smaller $N$ the critical temperature is larger, but the
vortex density is smaller. For instance, for $N=10$ we obtain $T_c\approx 30$ nK and
$n_v\approx 8\cdot 10^3$ cm$^{-2}$.

The vortex density is proportional to the dipole moment of the compound particle and
independent of its mass. Therefore the transition from the superfluid gas of chains to
$N$ uncoupled superfluid 2D gases should be accompanied with a strong decrease in the
vortex density. Such a transition can take place if one increases the distance between
the traps in the stack or decreases the electric field that aligns dipoles along the
$z$-axis.

At low density the statistics of individual molecules is not important and Fermi
molecules at even $N$ binds into Bose chains as well.  At larger densities Fermi systems
may demonstrate some specific features. It was shown in \cite{17} that the ground state
of  a Fermi gas of dipolar particles in the multilayer system is a dimerized superfluid,
with the Cooper pairing only between every other layer. The dimerized superfluid
\cite{17} is a variant of the Bardin-Cooper-Schrieffer (BCS) state \cite{21}.
 BCS state corresponds to the weak coupling  limit. The weak coupling limit can be also understood as the high density limit in a sense that the size of the Cooper pair is much larger than the average distance between the pairs.
   In the strong coupling limit the self-consistence equation for the BCS order parameter is reduced to the Schroedinger equation for the pair of particles \cite{20}.

The situation in multilayers is more complicate. The BCS state is the state with only the
two-body coupling.
The 4-body, 6-body etc. coupling is out of the BCS approximation and superfluid state of
chains cannot be described within the BCS approach. The pairing in multilayers is similar
to one in many-particle barionic systems, where the transition from the BCS paired
superfluid to a quartet Bose-Einstein condensation (BEC) occurs under decrease in density
\cite{22}.  By analogy with \cite{22} in multilayer dipolar Fermi gases a transition from
the dimerized superfluid to  the chain superfluid is expected. In nonuniform magnetic
field that induces vortices such a transition should be accompanied with a strong
increase of the vortex density.

 We would mention that multilayer ultracold polar gases are accessible now experimentally. In particular, in \cite{ex1} a
 multilayer ($N>20$) stack with polar $^{40}$K$^{87}$Rb molecules with the centre layer
having more than 2000 molecules and a peak density of $3.4\cdot 10^7$ cm$^{-2}$ at
$T=500$ nK was realized. A few-layer stack with polar $^{23}$Na$^{40}$K molecules at
$T=300$ nK was realized in \cite{ex2}. Also it was reported recently \cite{ex3} on a
realization of a degenerate three-dimensional Fermi gas  of $^{40}$K$^{87}$Rb molecules
at $T=50$ nK with  density $n\approx 2\cdot 10^{12}$ cm$^{-3}$ and total number of
molecules $3\cdot 10^4$. Therefore one can hope that degenerate multilayer dipole gases
will be obtained soon.
 Thus, current state of art in creating, cooling and trapping of dipole gases give expectations that systems with required parameters will be realized in the nearest future.

In conclusion, we have shown that crossed electric and magnetic fields can be considered
as a tool for the study of the phase transitions in multilayer dipolar  gases. In such
systems subjected to the electric field  that aligns the dipole moments perpendicular to
the layers  polar molecules bind  in long chains. At low
 temperature at which the gas of chains becomes the superfluid one nonuniform magnetic field with nonzero two-dimensional divergence
 may generate quantum vortices in a gas of chains.
The disappearance of the vortex pattern under variation of the parameters of the system
can be a signature of the initial presence of chains which then undergo dissociation.


\end{document}